# Conceptual Modeling and Classification of Events

**Sabah Al-Fedaghi** *

*Computer Engineering Department*
*Kuwait University*
*Kuwait*
**salfedaghi@yahoo.com, sabah.alfedaghi@ku.edu.kw**

*Abstract–*This paper is a sequel to an evolving research project on a diagrammatic methodology called thinging machine (TM). Initially, it was proposed as a base for conceptual modelling (e.g., conceptual UML) in areas such as requirement engineering. Conceptual modelling involves a high-level representation of a real-world system that integrates various components to refine it into a more concrete (computer) executable form. Methodologies for developing conceptual models have revealed that using well-defined ontological notations and well-structured processes is crucial in enhancing the efficiency of building (software) systems. The TM project has progressed into a more comprehensive approach by applying it in several research areas and expanding its theoretical and ontological foundation. Accordingly, the first part of the paper involves enhancing some TM aspects related to structuring events in existence, such as absent events. The second part of the paper focuses on how to classify *events* (e.g., typically named achievements, states, activities, and performances) and the kinds of relationships that can be recognised among events. The notion of events has occupied a central role in modelling. It influences computer science and such diverse disciplines as linguistics, probability theory, artificial intelligence, physics, philosophy and history. In TM, an event is defined as the so-called thimac (*thing/machine*) with a time breath (time subthimac) that infuses dynamism into the static description of the thimac called a *region*. A region is a diagrammatic specification based on five generic actions: *create*, *process*, *release*, *transfer* and *receive*. The results of this research provide (a) an enrichment of conceptual modelling, especially concerning varieties of existence, e.g., absent events of negative propositions, and (b) a proposal that instead of semantic categorisations of events, it is possible to develop a new type of classification based on graphs grounded on the TM model diagrams.

*Index Terms –* Conceptual model, diagram, structure, events, absent events, event classification, thinging machines

## I. INTRODUCTION

This paper is a sequel in evolving research about a diagrammatic methodology called thinging machine (TM) that have been proposed to be used as a base for conceptual modelling that serves in such areas as requirement engineering (e.g., on or above the level of UML as a conceptual modelling language). The TM project has evolved into a more comprehensive approach by applying it in other research areas and expanding its theoretical and ontological foundation.

Conceptual modelling involves developing a high-level representation of real-world systems, covering components and behaviours to be refined into a more concrete executable model. Outlining methodologies to build conceptual models involves difficulties related to modelling entity types, generalisation hierarchies, relationship types, attributes, and cardinalities [1].

------------------------

The evolution of newer techniques in conceptual modelling has revealed that using well-defined modelling notations and following clear processes improves effectiveness.

### A. Thinging Machines

The research in this paper contains two parts. In the first part, new aspects of thinging machines are further developed, which are related to structuring reality into two levels, with a potentiality/actuality scheme adopting an idea that goes back to the Stoic modes of reality.

The TM model provides us with an ontological representation of reality. It represents the entities that '*there are*' in the targeted domain and processes utilising one notion, *thimac* (*thing/machine*), which is acronymised as TM. Generally, a thimac (e.g., a chair, river, etc.) is conceptualised as a single thing/object and simultaneously a process/machine. It is represented as a diagram (called region) with five generic actions: create, process, release, transfer and receive. A TM event is constructed from a region and time.

### B. Event Classification

The second part of the paper focuses on foundational matters in TM, including the classification of events and the kinds of relationships that can be recognised among events. The notion of events has occupied a central role in modelling and influenced computer science and philosophy; therefore, considerable gain can be achieved by further examining this notion. An event plays a prominent role in various areas of philosophy, from metaphysics to the philosophy of action and mind, as well as in such diverse disciplines as linguistics, literary theory, probability theory, artificial intelligence, physics, and history. This plethora of concerns and applications indicates the prima facie centrality of the notion of an event in conceptual scheme. [2]

In software engineering, the notion of an event is an essential concept in an events log or audit trail, which is an ordered sequence of occurrences containing evidence of the execution of a process by users, systems, or other entities [3]. Event classification refers to defining the characteristics of a system's various sets of events and grouping them into logical categories to facilitate security objectives. In computer networks, monitoring of events includes Network traffic (packets) attempting to access the host, Log-in activity on the networking and any changes to the files.

Whitehead [4] affirmed that everything is an event. The world is made of events and nothing but events. Even a solid object is an event, a multiplicity, and a series of events. In a more recent discussion of the notion of event, according to [5], "Events are objects in the manifold of the three dimensions of space plus the one dimension of time […] the distinction between what we call "events" and what we call "objects" in everyday language marks the presence of interesting temporal structure; formally they are the same."

### C. Sections

Section 2 introduces the TM model with many enhancements. Section 3 supplements this with an entire case study. Section 4 elaborates on the nature of the TM model. The second part of the paper which examines the classification of events, is in sections 5 and 6,.

## II. TM MODELING

This section includes new clarification of TM materials discussed in previous papers (e.g., [6-9]).

### A. General Outlook

The TM model provides us with an ontological representation of reality. It represents the entities that 'there are' and processes in the targeted domain utilising one notion, which we call *thimac*. Fig. 1 shows the structure of thimac. As an initial example, Fig. 2 shows that the static TM model that involves four thimacs of *A statue is made of clay*.

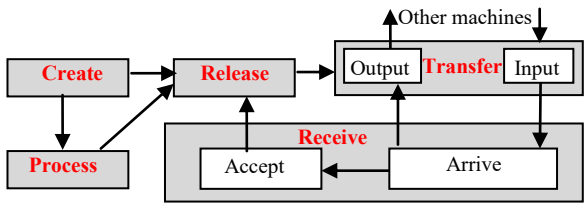

Fig. 1 Fundamental structure of a generic thinging machine.

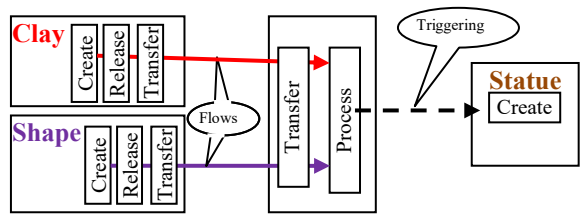

Fig. 2 Static TM model of *A statue is made out of clay*.

A generic thimac (A thimac that does not include any subthimac) is a gathering up into unity or synthesis of actions: create, process, release, transfer and receive. The thimac constituents are formed from the makeup of these actions. An action is a unit of actionality. A TM diagram like the one in Fig. 2 is called a *region* at the static modelling level. Fig. 3 shows a picture that outlines the two levels of the TM scheme. The synthesis of actions mentioned previously is also applied to *events* (at the dynamic level—see Fig. 3). Actions are synthesised just as a chemical compound is blended from atoms of different elements joined together by chemical bonds. Nevertheless, at the dynamic level, regions remain distinct from their events.

### B. Thimac Two Modes: Region and Event

Thimacs can be any object, property, trope, process, class, relation, or state. What distinguishes a thimac and a mere aggregate of constituents (things) is that a thimac has a structure that holds constituents together; while a mere aggregate of constituents does not. The TM structure includes the set-theoretic structure with order.

A thimac is a *structure* that involves *constituents* and their configuration (blueprint) and relates to other structures and actions of other thimacs in the domain of interest. A knife thimac blueprint, for instance, involves what a knife will do when brought into contact with all types of thimacs, e.g., butter and bread and so on, its involvement of world uses of a knife, etc., according to the aspect of interest in the targeted domain of modelling.

The most basic thimacs *have* characterisations/properties (subthimacs) and *act* through the five TM actions. These five actions are essential and sufficient for thing-ness and machine-ness (to be discussed next).

Additionally, a complex thimac is a thimac that interweaves with other thimacs. The world of thimacs, when viewed as representing reality, *characterises* what could be the case (potentiality – regions) but only *represents* what the case (existence – events) is. We conceive the TM to function as diagrams that provide ontological explanations standing for the modelled world of static regions and dynamic events.

In this conception, a region is a construct of what *might be* an event. Regions have the same sense of 'meanings' as they function as parts of events. They describe their theimacs components (subthimacs) and actions. Events are based on these configurations of regions and their *temporal* structure. As shown later, regions are represented as diagrams at the static level (e.g., Fig. 2), and events are superimposed on these diagrams to produce new diagrams representing the dynamic model.

### C. Realisation of Regions into Events

Even though it is not the focus of the explanation in this paper, it should be understood that realising a region in an event may involve other things besides time, such as energy and matter. Such things as energy and matter are thimacs that may be subthimas of regions, just as colour is a subthimac of a thing (Examples will be given later in this paper). While a region may have shape, size subthimacs, it may have also subthimacs of (type of matter), mass, hardness, etc. It has all its 'categorical traits' [10].

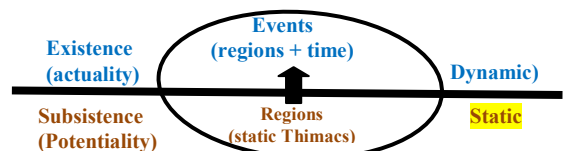

Fig. 3 Two levels TM modeling

## D. Thimac may be a Thing, or Process

A thimac may stand for a thing or a *process*. The difference between what thimacs do and what merely happens to them reflects a difference between the active and the passive thimac or the traditional notions of agents and patients. A thimac can have two roles: as an agent (that creates, processes, releases, transfers and receives) and as a patient (being processed, being released, being transferred, being received).

The notion of existence/subsistence (Fig. 3) is extended to process; a process takes a 'thing/object-like' form or may, in classical language, be 'the predicate of propositions. Generally, a TM thimac is suited to be a thing/machine, i.e., creates, processes, releases, transfers, and receives, in addition to being created, processed, released, transferred and received. Such a notion is applied also to *relations*. For example, *John is happy, and Mary is happy* form at the existence level two events and a (a whole) super event (corresponds to *and*), where the super event is more than its events constituents (i.e., it has its own *create*).

## E. Create as Minimal Actionality in Thimacs

*Create,* designate, or express the notion of a 'subsistence piece' (thimac) at the static level and an 'existence piece' (thimac) at the actuality level. *Create* (at the static level) is not a name or label of a thimac; the thimac static sense requires an assertion (judgement) in an explicit verbal form (an action that provides progressing in time) to install a thimac in the category of the modelled domain. *Create* simply indicates that the thimac *is* a thimac. Hence, *Aristotle subsists,* or *Aristotle exists* is replaced by the thimac *[Aristotle: create]* as a region or event, respectively. Aristotle in *[Aristotle: create]* is an event at the dynamic level if it is a thimac-based 'unique description' (Russelian term), which belong to nothing but *Aristotle*. Similarly, *[It is raining, create]* is a declaration of a thimac. At the static level, *[It is raining, create]* is a region, and it is an event when it is (actually) raining (e.g., now). Very complex processes may need no name, but its *creation* is necessary to register it as an ontological unit.

Actuality (see Fig. 3) includes all constituents in the event besides 'pieces of existence', called exicons [6]. None of the constituents are actual without them (i.e., exicons). The thimac must have a minimal actionality of 'create'. Every thimac must have the action *create* but *create* (and other TM actions) is not a thimac. In classical terminology, we can say that the thimac is the subject of the predicate since (logically) every proposition must have a subject and a predicate. In some diagrams in this paper, for simplicity's sake, we may ignore *create* in a thimac under the assumption that the outlining diagram rectangle of the thimac is sufficient for its declaration as a thimac in the TM model.

## F. The Thimac: Thing and Machine

The thimac has a dual nature of a thing and a machine. It is a machine that acts on things. The machine "collapses" in a thing when it is acted on by machines (see Fig. 4).

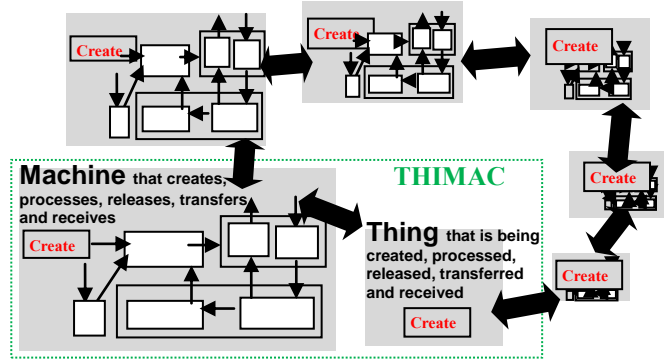

Fig. 4 The collapsing of thinging machine into a thing

A thing is an 'Irreducible exemplification" (whole) of a machine (composite of actions) that 'stays the same' when acted on.

## G. Actions

Thimac's actions, shown in Fig. 1, are described as follows.

1) *Arrive:* A thing arrives at a machine.
2) *Accept:* A thing enters the machine. For simplification, the arriving things are assumed to be *accepted* (see Fig. 1); therefore, *arrive* and *accept* combine actions into the **receive** action. Thus, the thing becomes inside the machine.
3) *Release:* A thing is ready for transfer outside the machine.
4) *Process:* A thing is changed, handled, and examined, but no new thimac is generated.
5) *Transfer:* A thing crosses a boundary as input or output from a machine.
6) *Create:* A thing is revealed in a machine, i.e., a thimac is registered as an ontological unit.

Additionally, the TM model includes *storage* (represented as a **cylinder** in the TM diagram) and *triggering* (denoted by a **dashed arrow**). Triggering transforms from one series of movements to another (e.g., electricity triggers heat generation).

Fig. 1 defines TM machines as their blueprint. This blueprint is a static 'form' at the potential level and becomes a dynamic event (a mutual fitting of region and time) at the dynamic level. Events cannot add to regions (the *whatness* of a thing) when transforming static regions into existence. This transformation presupposes 'knowing' existence in some other way, e.g., sensation.

The thimac *machine* executes five actions: *create*, *process*, *release*, *transfer,* and *receive*. Each static (outside time) action is a capacity to act and becomes a *generic* (has no sub-event) event when merged with time. A thing can be created, processed, released, transferred, or received. All things can flow (i.e., things that can be created, released, transferred, received, and processed). A thing flows from a machine through its *release* and *transfer* (output) to reach a second machine's transfer (input) to settle in its receive.

## III. EXAMPLE OF TM MODELLING

According to [11], propositional logic has a wide variety of applications in computer science. It is used in system specifications, circuit design, logical puzzles, and more. It can also translate English sentences into mathematical statements and vice versa.

*A.  Example I* (From [11]): Let *a, b, c,* and *d* represent the following sentences.
   *a: The computer is within the local network,*
   *b: The computer has a valid login ID,*
   *c: The computer is under the use of the administrator,* and
   *d: The Internet is accessible to the computer.*

So the complex sentence, *If the computer is within the local network or it is not within the local network but has a valid login ID or it is under the use of an administrator, then the Internet is accessible to the computer,* can be expressed as $(a \lor (\neg a \land b) \lor c) \rightarrow d$ [11]. Fig. 5 shows the corresponding TM model of the involved complex sentence. In this representation, the expression *The Internet is accessible to the computer* refers to two-way communication,

-   The computer sends something to the Internet, and,
-   The internet sends something to the computer.

### Static Model

The static TM model (Fig. 5) shows these two communication streams on the box's left and right sides, labeled 'local network'. Accordingly, the static model of Fig. 5 involves the following – for simplicity's sake, the figure consists of some redundancy.  Note that the circled numbers are just labels to explain different positions in the diagram.

**1.** Communication from the computer to the Internet
-   The computer sends a message requesting access to the Internet (circle 1 in Fig. 5).
-   The local network sends the computer request to the controlling module (2).
-   The message is processed to extract the computer ID (3).
-   The list of computers within the local network is retrieved from the local network (4).
-   The list of computers within the local network and the extracted computer ID are processed to determine whether the ID is in the list (5).
-   If the computer ID is in the list of computers within the local network (6), the computer message is released to the Internet (7 and 8).
-   *If the* computer ID is not in the list (9) of computers within the local network, retrieve and process the computer ID (10 and 11).

-   If the computer ID is valid AND (11) the computer ID is not in the list (9) of IDs which are within the local network [This models $(\neg a \land b)$], then release the computer message to the Internet (12). Note that the horizontal thick black line (12) simplifies the logical *OR* of the two signals.
-   Retrieve *the under administrative use* (13), and if the computer is under administrative use [*This model's proposition c*]**,** then release the computer message to the internet (14).

**2.** The Internet sends a message to the computer
-   The message arrives at the control module (15).
-   The Internet message is processed to extract the destined computer ID (16).
-   The list of computers within the local network is retrieved from the local network (17).
-    The extracted computer ID and the list are processed (18)
-   If the extracted computer ID is in the list (19), the Internet message is released to the computer (20).
-   If the extracted computer ID is **not** in the list (21), retrieve and process it (10 and 11).
-   If the computer ID is valid AND (11) the computer ID is **not** in the list (21) of computers that are within the local network, then release the Internet message to the computer (22)
-    Retrieve the *under administrative use* (23), and if the computer is under administrative use (24), then release the Internet message to the computer (25).

### Dynamic model

Fig. 6 shows the TM events model where each event is represented by its region (subdiagram of the static model). Note that the events corresponding to proposition a are represented by the events *$a_1$* (*The computer is within the local network* with respect to outbound communication) and *$a_2$* (*The computer is within the local network* with respect to inbound communication). Similarly, $c_1$ and $c_2$ represent the proposition c.  Accordingly, the events in the model are listed as follows (not in the chronology of events order).

**From the computer to the Internet**:
$e_1$: The computer generates a message to the Internet.
$e_2$: The computer ID is extracted and compared with the list of computers within the local network.
$a_1$: The computer ID is within the local network.
$\neg a_1$: The computer ID is not within the local network.
$b$: The computer ID is valid.
$c_1$: The computer is under the use of an administrator.
$b \lor \neg a_1$: The computer ID is valid, AND the computer is not in the list.
$d_1$: The computer message is released to the Internet.

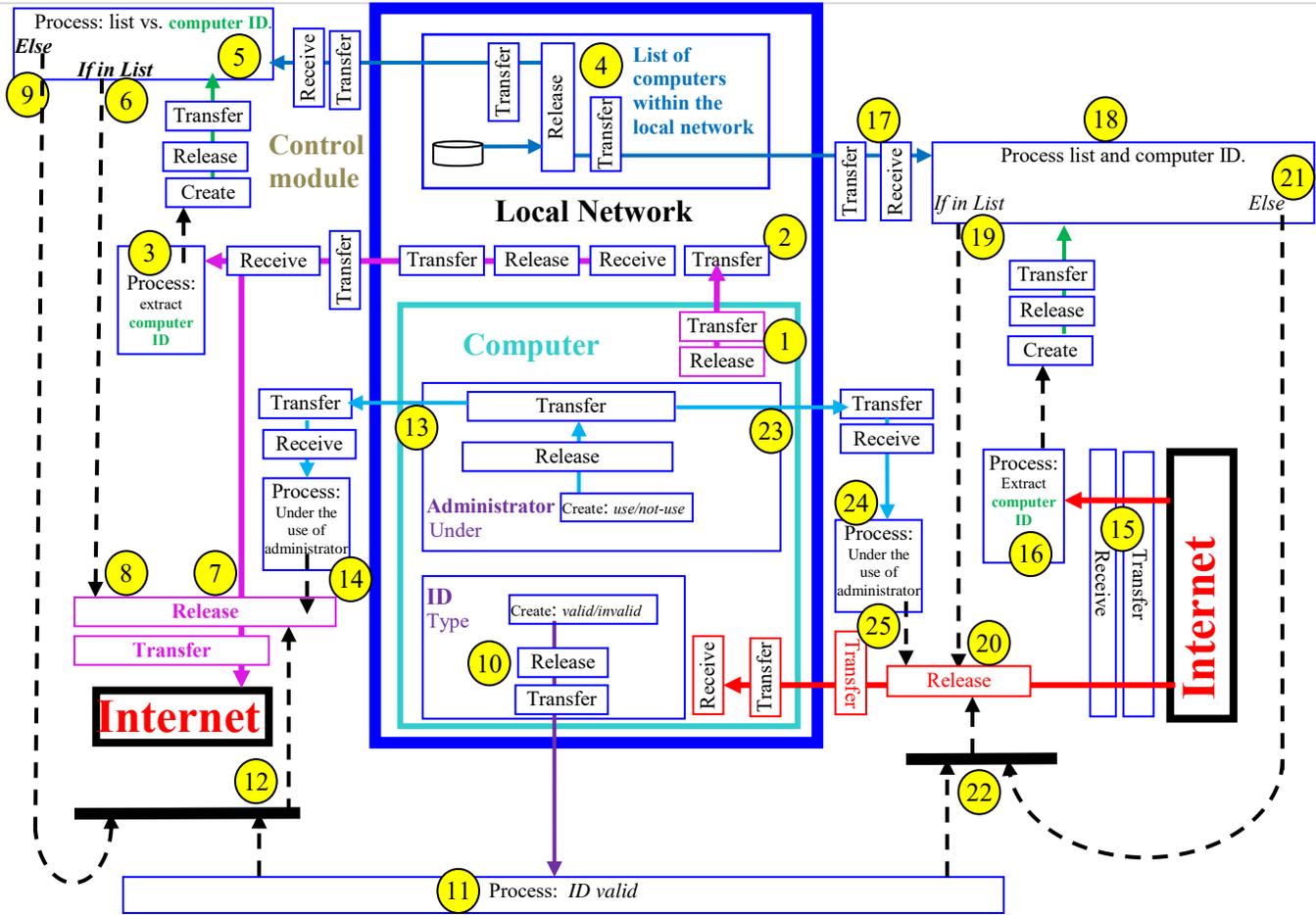

Fig. 5 The static TM model of $(a \lor (\neg a \land b) \lor c) \to d$.

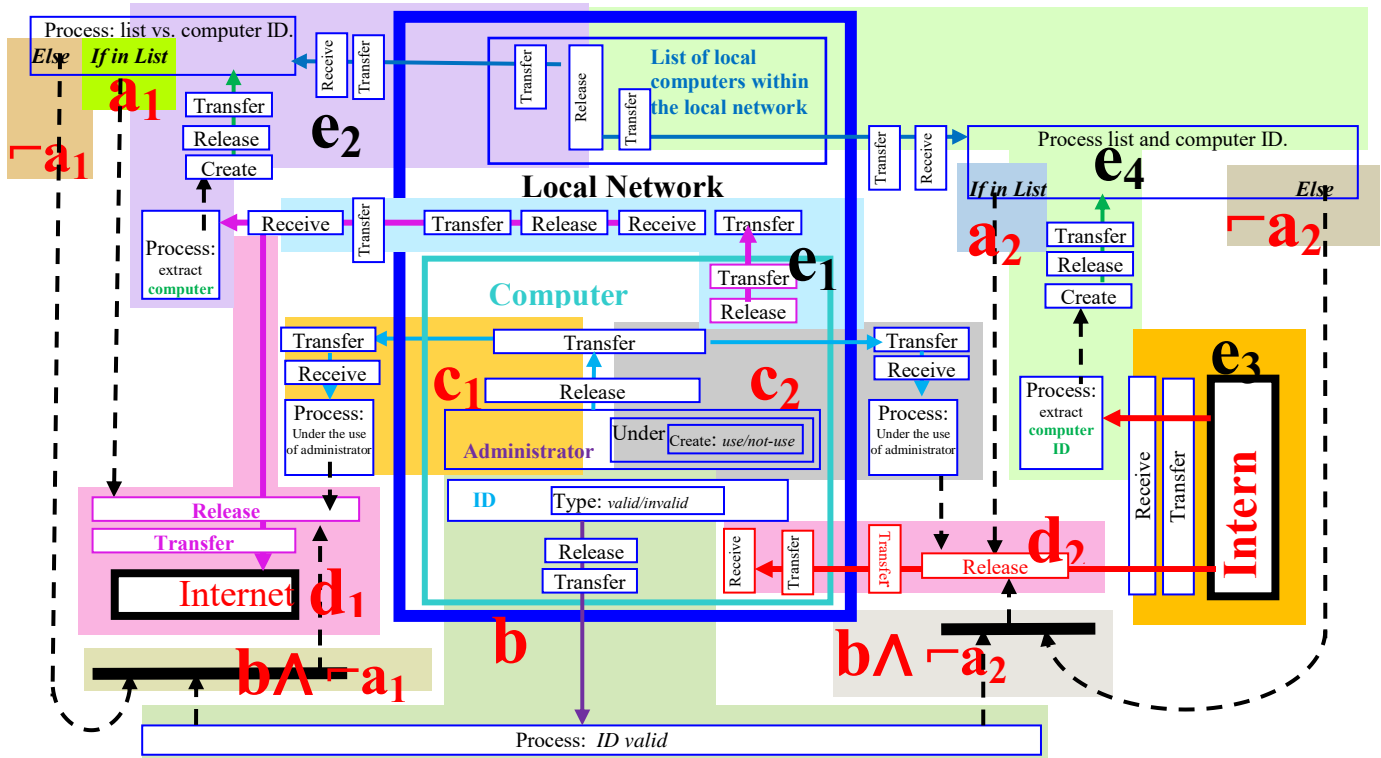

Fig. 6 The dynamic TM model of $(a \lor (\neg a \land b) \lor c) \to d$.

**From the Internet to the computer**

$e_3$: An Internet message arrives destined for the computer.

$e_4$: The destined computer ID is extracted and compared with the list of local computers.

$a_2$: The computer ID is in the list

$\neg a_2$: The computer ID is not within the list of the local network.

$b$: The computer ID is valid.

$c_2$: The computer is under the use of administrator.

$d_2$: The Internet message is released to the computer.

Fig. 7 shows this model's chronology of events. The TM model furnishes the underlining semantics of $(a \lor (\neg a \land b) \lor c) \rightarrow d$.

### B. *Example* II

As we saw in example I, TM diagrams involve a static model, a dynamic model, and a chronology of events. This section illustrates the fourth and last diagram, the timing model.

In TM, actions have two modes: (timeless) static and dynamic. To illustrate the time aspect of an action, consider what Aristotle says: "For there is no difference between saying that a *man walks* and saying that *a man is walking*" [12]. However, the corresponding events in TM differ, as shown in Fig. 8. Fig. 8 shows the timing diagram in TM modelling. It specifies the extension of event durations. Thus, a man as an event extends in time more than walking.

### IV. ELABORATION ON TM REGIONS AND EVENTS

Section 2 furnished the basic TM modeling concepts necessary for practicing it, as demonstrated in section 3. This section provides notions for further studies to understand the model better. Specifically, two topics are discussed,

- Exploring the meaning of the relationship between regions and events.
- When describing the targeted domain, elaborate on ontological representing of absent events that model negative expressions (e.g., in the Internet example in section 3: *The computer is **not** within the local network*).

### A. *Event and region are inseparable but not the same.*

As described in [13], existence (events) and subsistence (regions) are like a double-image impression where the two are inseparable in reality but not the same. There is no region of an event (i.e., inside the event) that *never* exists. Likewise, there is no event of a region that never subsists as a region. Regions require no more validation of their subsistence than does time. Think of events without their regions (static constituents). Even a 'pure' event, say electrical energy, necessitates a non-dynamic picture of electrons and movement trajectory: release, transfer, and receive from one level to another. In all such examples, the event and region are inseparable in reality but not the same because of the static/dynamic divide.

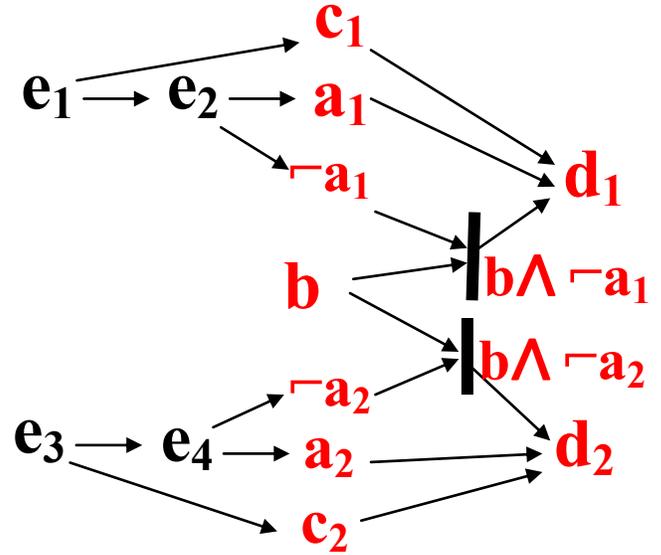

Fig. 7 The chronology of events of $(a \lor (\neg a \land b) \lor c) \rightarrow d$.

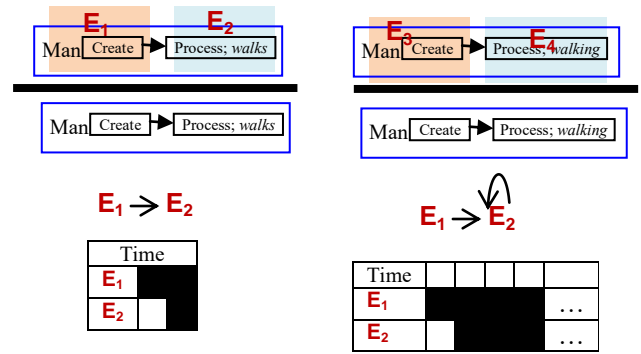

Fig. 8 TM representation of *Man walks* (left) and *Man is walking* (Right)

When we see an event, we simultaneously perceive (may need to exert some efforts, e.g., as in Rubin's vase) its region — not necessary in the TM thimac diagrammatic form. Regions are residents in the events. They are perceived in no other way than as integral parts of events. Events and their regions are perceived holistically. They are the way of grasping the *existence* of a thing as a distinct phenomenon from the thing itself *out there*.

When I perceive an event, I perceive its region, where I recognize its configuration (structure). Perceiving regions are analogous to the cognition of Plato's forms but without introducing the notion of the soul's encounter with forms. A region exists — to imitate Plato's description, is locked in its event "like an oyster in its shell'' — and is attainable by our senses (observed).

According to [14], "By analogy, when an animal perceives the white of the sun, they assimilate the chromatic form of the sun but none of its matter. Moreover, just as it is the form of a ring, and not in gold or bronze, that produces the sealed impression, its distinctive shape, it is the whiteness of the sun, and not its matter, that produces the sensory impression, the perceptual experience of the white of the sun." Additionally, to quote Aristotle,

When a man has in his mind a good thick slab of wax, smooth and kneaded to the right consistency, and the impressions that come through the senses are stamped on these tables of the 'heart'—Homer's words hint at the mind's likeness to wax—then the *imprints* are clear and deep enough to last a long time. [14] (Italic added)

TM can model such processes by replacing forms with regions, as illustrated in Fig. 9. In Fig. 9, the region (of cloud) perception is the assimilation of the region without the perceived (physical) event. In such a phenomenon, we can say that regions pass (flow) through events. In this case, for example, *colour* (see Fig. 9) is a sub-theme of the perceived object, thus a sub-region that is printed in the object event in a manner analogous to Plato's metaphor of wax's reception of the impression sealed by a signet ring. The *region* is transferred from one realization (event) to another realization (event) that differs, for example, in (physical) 'matter' that composes them.

Fig. 10 shows the corresponding events of the dynamic model, and Fig. 11 shows the chronology of events. The backward arrows in Fig. 11 indicate continuous flows of regions. Of course, 'experiencing' regions does not imply a diagrammatic representation. It is just a conceptual model of a "hypothetical structure that we describe and investigate in the hope of using it to understand some more complex, real-world 'target' system or domain" [15].

In Heidegger's language, this modeling domain is the unconcealment of a procedurally prior concealment that results in the target domain coming to be 'lit up.' This unconcealment can be thought of as "the dust on the floor is not meaningfully present to me at all until I *uncover* it when I start (to think about) cleaning" (Italic added) [16]. Unconcealment "is an event" [17] understood to be the self-manifestation of a thing (thimac) and the disclosure of constitutive ontological structure (region) and function of an event.

In traditional language, sensible color (region) exists in events (external particulars). Note that the light rays (Fig. 9 and 10) from a perceived object that reach our eyes form an event thimac. This thimac carries the imprint of the color region to the eye, which is, in turn, an (event) thimac. In present-day physics, "descriptions of this kind [color names] are nothing but abbreviations, since they presuppose wave theory and since the color names can be translated into expressions of this theory (i.e., rates of oscillation)" [18].

Regions are abstracted as 'diagrammatic things' that are not physical. A region is the 'form' of an event projected over the static world. We usually don't explicitly take notice of regions, or they can hardly be noticed when we figure out the characteristics of the event, e.g., color, and simultaneously accept its existence as the shadow of the wholeness of an event.

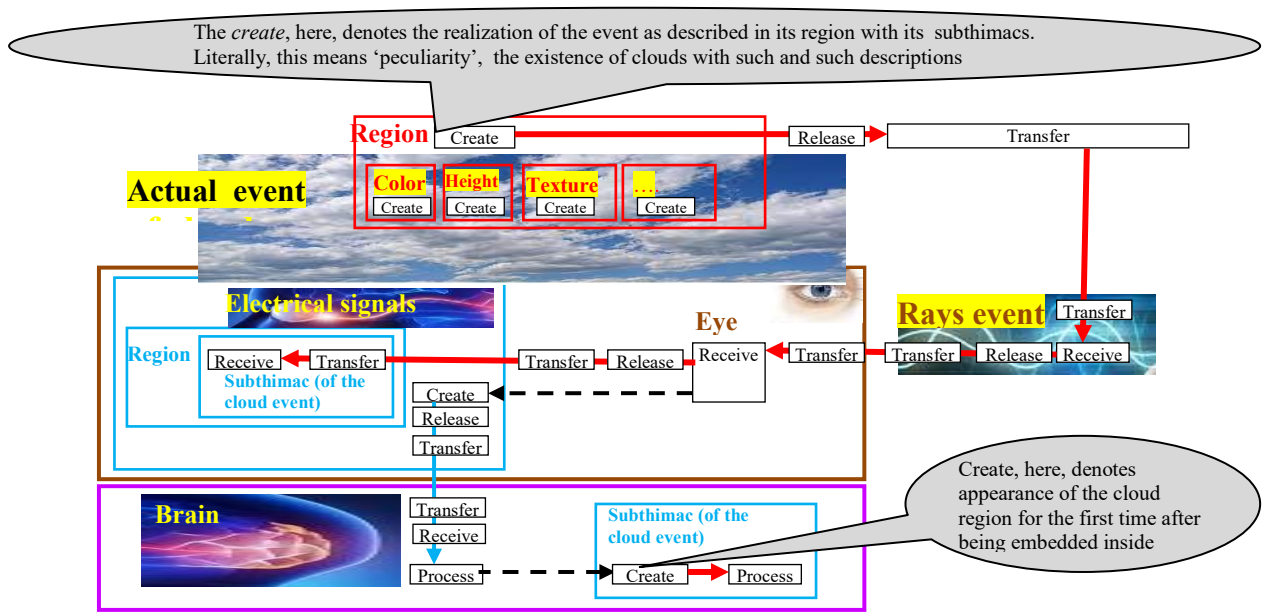

Fig. 9 A simplified illustration of the flow of event region

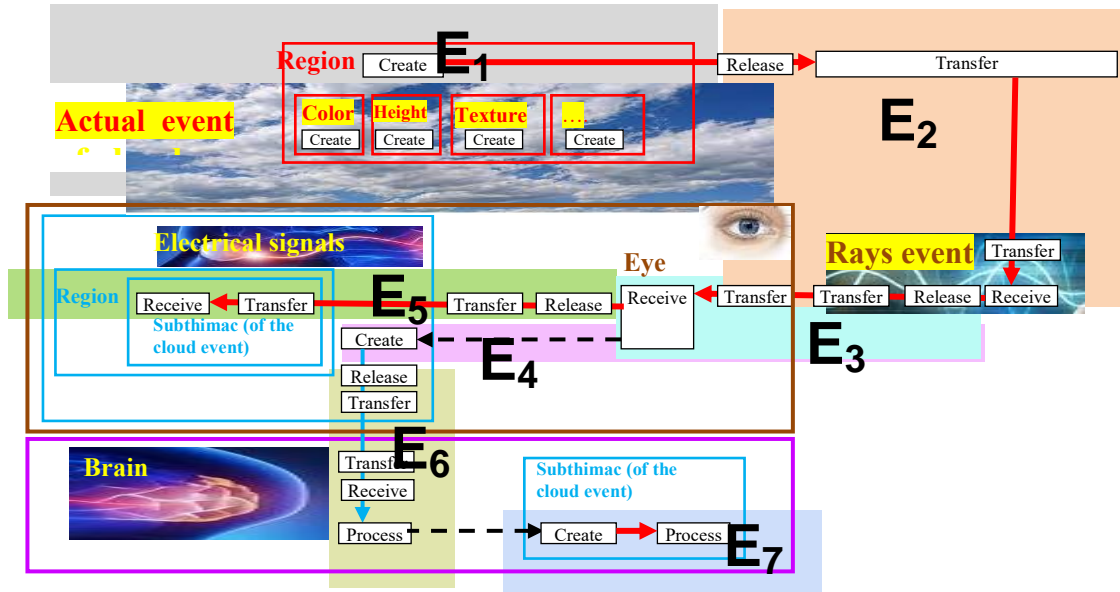

Fig. 10 Dynamic model

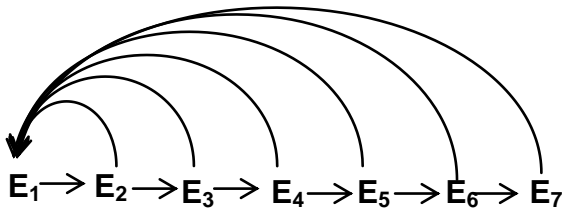

Fig. 11 Events model

## B. Absent Events and Negative expressions

The notion of *absent* events is defined as events that model negative propositions (See [6])] and is important in TM. A region of an absent event refers to a certain absence in contrast to the *present* existence exemplified by positive propositions describing the targeted domain. Previously, absent and present event/propositions were given in the Internet example in section 3 without directly mentioning this distinction.

*John is not in the room*, indicating the absence of sub-event *John*. That is, a part of the larger region *John is not in the room* (*John*) has been realized in absence. The event here is like a piece of Swiss cheese; we don't ask where the cheese isn't in the piece. Note that a region as a part of an event (existence) is not a 'hard' state of affairs but a fluid dynamic cluster of action(s). Thus, in a dynamic generic region (generic event – e.g., create, process, release, transfer and receive), we cannot say it (existence) is at a certain 'place' at any point in time of a single generic action (See [6] — discussion of Zeno's puzzles). The nature of change in generic events can be illustrated as a boundary between two generic events that overlap where the first generic event disappears and the successor event appears. Accordingly, each action has three breakpoints: the action (e.g., process) 'flows' from an initial state, a vibrant change to a next state (See [6] — graph of transformation between states). A state here refers to an action in existence.

For example, *False gold* (negative fact) is realizable as an absent (gold) event. It exists, but 'gold-ness' is absent, as shown in Fig. 12.

Every event, present or absent, has a region. An 'event without a region' does not exist just as, according to *folklore*, vampires do not exist since they cannot have images in mirrors. Events (present and absent) occupy 'existence slots' (called previously in this paper *lexicons*) that resonate (integrate) regions since event features (presence or absence) are perceivable (experienced).

Imagine standing in the sun, and then a shadow emerges from you. The shadow defines the precise shape, size, etc. It is the 'actual region' of our presence in the sun (event). It 'exists,' and its *absence* of actuality means it is a constituent of events but not an event (subevent) itself.

Regions do not change (e.g., lose/gain an action in their diagrammatic description) when in events like the changeless shadow of a rotating sphere. Thus, regions are *subsisting* on real things inside perceptible events. Regions are shadows of events. They cannot change, and they do not have mass or energy. Hence, they cannot have some basic features of the events they are shadows.

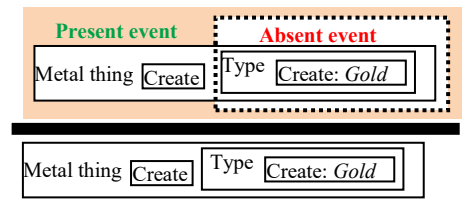

Fig. 12 TM representation of *False Gold*

A region in existence is "absently present" [19] or, in TM, an *absent event*. In other words, a negative region corresponds to an *absent* event instead of a *present* event (of a positive region). An absent and present event may have the same region, just as red and golden spheres may have the same shape. Note that such a negation as ⌐*(round square)* is not included here because *it* is not realizable, i.e., it does not have the event *round square*. In TM, a negative region should negate a positive region to correspond to an absent event. Since the proposition *round square exists* has no realization, its negation is excluded from the existence level (it is not an absent event).

Accordingly, consider the statement: *there is a square circle* of which it is true that *there is no such a square circle*. Then, from the TM point of view,

- 'There is a square circle' means a region without a corresponding present event.
- 'of which it is **true**' means its corresponding event.
- 'there is no such a square circle' means there is no absent event.

Fig. 13 illustrates this situation. As mentioned, we model only actualized regions (realized or possible) in TM modelling.

The issue discussed here concerns the truth of negative propositions and whether to accept or reject their existence. In TM, a negative proposition is a negative region. Assuming the realization of its positive version, then the realization of negative region is an absent event. Here, implicit negation, such as opposite propositions, does not involve negativity, as shown in Fig. 14.

In TM, we split events in existence into events in presence and events in absence. A negative region subsists, and its realization (existence) is its absent event. Assume we presuppose the law of excluded middle at the static level. In that case, a region is either positive or negative, and its event is present or absent. This type of thought may be close to Wittgenstein (Tractatus), as he distinguishes between "the totality of states of affairs that obtain" [TM subsistence] from actuality, which, he tells us, is "the obtaining and non-obtaining of states of affairs" [TM existence].

### C. Metaphysics of Absent Event

Existence includes (maybe there are more to uncover in further research, e.g., mental events) two modes: in *presence* and *absence*. In *present existence*, the existent is a 'thing' with unity and performance of action (creates, process, etc.). The *negation* of subsistence is an absence at the existence level where the region is an unspoken 'language', i.e., the region diagram is not 'articulated', i.e., lack of capability to act (see [6]). This line of thought is analogous to Wittgenstein's thought [Tractatus], "The existence of atomic facts we also call a positive fact, their non-existence a negative fact."

Accordingly, we find in the following two cases, a region can feature a lack of capability to perform an action or lack of effect.

- A region at the *static* level region, and also,
- The region of an absent event at the *existence* level.

Note that the effect is defined here regarding the actions create, process, release, transfer and receive

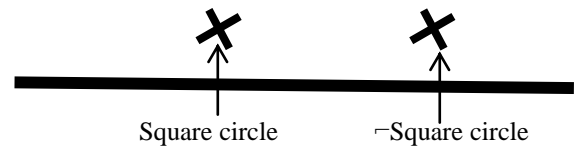

Fig. 13 An eventless region has no absent event of its negation

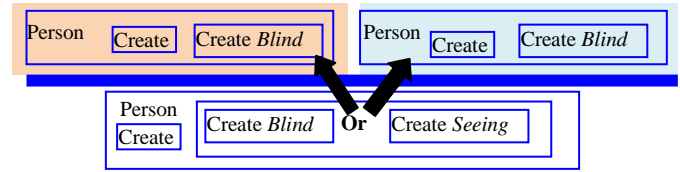

Fig. 14 TM representation of opposite properties

In the abovementioned region of a negative proposition, the region inside an event *exists* but is simultaneously *absent*. Accordingly, the notion of something existing is applied to performing actions (effectiveness) and to things that are absent events, and their regions have no power of action. In the following discussion, we further illustrate and describe the notion of absence at the existence level.

These absent things (as a subthimacs of events) can be described in analogy to an *image* (say, a man) in a mirror. The image exists because it (viewed as a patient) has a **presence** in terms of behaviour or can act upon the visual power and *be seen*. The image engages with the outside regarding what it is: *a displayer*. But the image subregion (e.g., eyes, nose, eyebrows, etc.) is mute, passive, **absent** and cannot see, smell, move, etc., by itself. In Heidegger's language, the eyes as part of the image inside the mirror lack their ontic nature, i.e., availability and usefulness for seeing while preserving their ontological nature (existence).

Even if the eyes, say produces tears inside the image, such action is an absent sub-event because the tears do not have their properties (sub-thimacs) such as wetness. In Heidegger's language, the eyes conceal themselves – "the **absence** of the conditions under which the entity in question could manifest itself in being" [17].

The *image* appears meaningful (carries information, unconcealed), but the eyes, nose, and eyebrows show up uninformatively or could not show up as completely what they are. They show existence (create) but not behaviour (process, release, transfer and release). If you block the image by a piece of paper, it seizes to function (displaying). Still, if you block only the eyes, they do not seize their functionality because the eyes in the image already lack the seeing functionality.

The image has the power of effect (e.g., it 'displays itself: create, release, to the outside (of the mirror). On the other hand, the eyes are displayed as a part of the image, but the display is not part of their 'functionality. By itself, it cannot transfer, receive, or send 'data' to the 'brain' of the image in it.

Using Heidegger's language, in our example, the eyes in the mirror unconcealed themselves as part of the man's image and concealed themselves as functional eyes. They lose their usefulness as seeing instruments and become *present-at-hand*.

Note that an image is in the mirror, and the region of the absent event is in the image.

An absent event *exists,* but its region cannot act. For example, if *John is not in the room, it* is an event, i.e., an existing thing (we sense it). Inside, *John is not in the room* (now); we recognise the room as an existing event but also recognise that *John* is in the image of the room. *John* is an *absent* subevent that cannot realise action — e.g., does not move out of the mirror (see Fig. 15). Another way of explaining this is that *John is not in the room* (now), expresses that *Room* is a present event and inside it the absent event *John*.

Now, suppose Mary replaces (takes place instead of) the mirror and says, "*John is not in the room.*" This is illustrated in Fig. 16. Both the *Room* and *John* are events (exist). The room is perceived and conceived, and John exists from previous experience but is conceived as a mental image that is not causally efficacious.

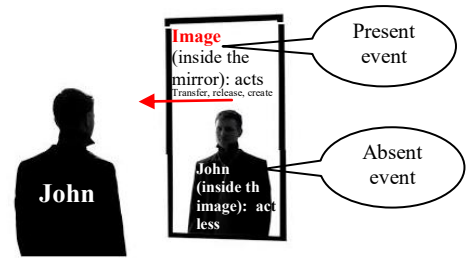

Fig. 15 Illustration of present and absent event using the analogy of mirroring

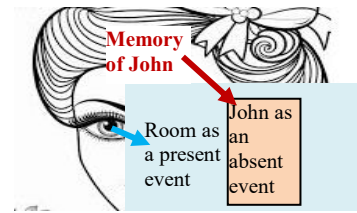

Fig. 16 Mary instead of the mirror has the room as an event and John as an absent subevent.

## V. A Glimpse of Events Per Se

According to [5], contemporary analytic philosophy has made several proposals about characterising events.

- Events should be represented as primitives, just as are objects.
- Events can be reduced to talk about facts.
- Events exemplify a property by some substance (particular) x at time t.
- Events are objects in the manifold of the three dimensions of space plus the one dimension of time.
- Events are gestalt in the stream of activity that flows through time.

[5] concludes, "Observers recognise *objects* by their distinctive shapes, colours, textures, tactile properties, and motion. They recognise *events* based on these features of their component objects and configurations of objects, but also based on their temporal structure." (Italic added). This description can be considered a 'simple' description of *regions* and *events* in TM.

Events mean changes. They are "gestalts in the stream of activity that flows through time" ([5] – see its source). An event does not exist, rather; it is a 'slot' of existence (called exicon in [20]). An exicon is an exemplification of existence, thus we use 'being event' as synonymous with 'existence' (regions and exicons fusion). It has a mode of existence that is absent vs. present events or *single* existence vs. universal existence. Events imply change either inside actions or among (across) them. Change may occur in *create,* indicating a change in existence (e.g., mere continuation), change in the *process* means internal change (e.g., colour), and change in other action implies a change in 'position' in the thimacs web (texture).

Heidegger argued that "no ontology can be sufficient without assigning being an eventual nature itself [...] The ruptural form of an event comes to play a central part in the ontological structures of time, ground, truth, language, history, community, the psyche, and so on" [21]. Being is described as evental in nature [21]. The event can be defined as "a fundamental, nonperceptual happening that form the basic building block of the Real as such. Everything, including people and things, can be constructed from events" [22]. While the thimac is more primordial than events, all actual things (e.g., elementary particles like electrons and quarks) are events. Events (hence, their thimacs) form event assemblages, higher-order event assemblages, and so on [22].

## VI. Events Classification

According to [23], event classification aims to identify a small number of event types into which all events can be classified. This classification has pinpointed basic features of events and has been used in virtually all investigations into event representation.

Aristotle distinguished between states and events, between events' terminal points, and between those that are ongoing with no definite terminus. Aristotle proposed three event types [23],

- An *actuality* that expresses "the existence of the thing" is interpreted to be a state.
- A *movement* is an incomplete process where an event lacks an inherent terminus.
- An *action* is a process with an inherent end.

In many modern studies, the Vendler scheme [24] has provided a four-fold distinction of verb types: *activities*, *accomplishments*, *achievements*, and *states*. In TM, the thimac is strictly verb-based since its constituents are actions. The central ontological value of verbs recognised by Aristotle,

> Every statement-making sentence must contain a verb or an inflection of a verb. For even the definition of man is not yet a statement-making sentence – unless 'is' or 'will be' or 'was' or something of this sort is added … A verb is what additionally *signifies time*, no part of it being significant separately; and it is a *sign of things* said of something else… So, every affirmation will contain either a name and a verb or an indefinite name and a verb. Without a verb, there will be no affirmation or negation. 'Is', 'will be', 'was', 'becomes', and like are verbs according to *what we laid down* since they additionally signify time (This quote is taken from [12]). (Italic added)

*A. Samples*

The Vendler scheme [24] can be grasped by reflecting on some examples that Vendler cited under each category (we assume all examples are events, i.e., timed).

**Activities**: As in verbs such as *run* (around, all over), *walk* (and walk), *swim* (along. past), and *push* (a cart) [12]. Fig. 17 shows the TM model of *run (around, all over)*. It expresses the following: a *thing* inside *all over* (e.g., in a stadium) exists, and then the *thing* flows *all over* to perform the process of running. The dynamic model expresses the chronology of events, and the timing table in Fig. 17 shows the implicit assumption that the thing exists before and after the run. There is no explicit mention that the run finishes since this is understood.

**Accomplishments** include *running* a mile, *painting* a picture, *growing up, and recovering* from illness. Fig. 18 shows the TM model of *running a mile*. It expresses two thimacs: *a thing* and *a distance (*one mile) exist, and the thing flows (processes itself) to run. The timing model is obvious, so it is unnecessary to repeat it. Distance is a 'universal' subevent, so it does not exist alone, as discussed previously.

Accomplishment is said to have 'an inherent end' by the agent, e.g., reaching one mile of running. From the TM point of view, there seems to be no fundamental structural (graphwise) difference between these cases (activities and accomplishments) to categorise them into different classes. The difference is in the implicit specification of the end of $E_2$ (in Figs. 17 and 18) when it is limited to one mile. In Fig. 17, the mere event of run (regardless of length) is an accomplishment as in the case of a person with a physical disability.

Some other researchers adopted these Aristotle's classes, but labelled them as: *states*, *activities* (actions with no terminus), and *performances* (actions with terminal state) [23]. For example, activity: *Terry is running* and performance: *Terry is building a house*. According to [23], "a crucial difference between activities and performances turns out to be one of *delimitation*. A delimited event is one that has an inherent or natural end".

Fig. 19 shows the TM representation of *Terry is running* (top) and *Terry is building a house* (bottom). In Fig. 19 (bottom), $E_1$ denotes that Terry exists and his house exists since the 'initiation' (not in a physical form) of the house (e.g., blueprint, location, etc.) as an unfinished project ($E_2$). At a particular time, the house comes under the active role of Terry ($E_3$) to be processed/built ($E_4$). During this course (becoming) of the building, the house is processed to determine whether the building has been completed ($E_5$). If the building has not finished, the building *continues* (backward arrow); otherwise, the house state reaches a finished product ($E_6$). Note that the *house* was originally a pure (no content) *concept*. The concept makes its entry into a *region* of a thimac (e.g., cost, plan, loan to build, contractor, etc.) to progress in detail and become an *event* in existence. The original concept may say to be an abstract, but the region and event are real things.

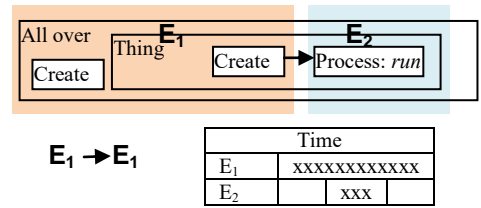

Fig. 17 TM model of *Run (around, all over)*

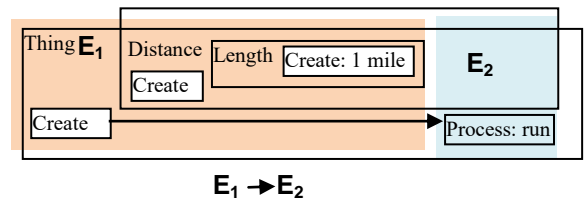

Fig. 18 TM model of *Run a mile*

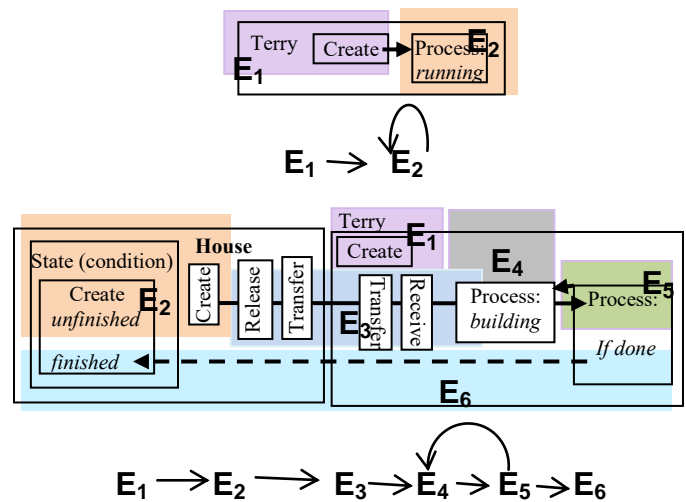

Fig. 19 TM representation of *Terry is running* (top) and *Terry is building a house*

Additional examples are given in [23], including,

**Activities**
(a) *Terry walked for an hour.*

*From* the perspective, Terry walking *for an hour* is an event since it includes past tense. Additionally, this event consists of a subthimac of its duration of walking: one hour. Fig 20 shows the TM representation of two events, E₁ and E₂. E₁ refers to *Terry's* existence, which precedes the walk. E₂ represents the event of the walk in the past tense (of the event not region) and the duration of one hour. The time diagram in the figure denotes the totality of this event: Terry exists then the event of walking in the past with one hour duration. The three components capture all the semantics of *Terry walked for an hour*.

(b) *Terry is driving the car.*

From the TM's point of view, *Terry driving the car* is not an event. If we assume that *Terry is driving the car now*, then its TM representation is shown in Fig. 21. We ignore the time 'now' since this timing element is illustrated in the previous example.

*Terry is driving the car now*, which indicates repeatability (reflexive arrow) in the duration *now*, doing the process (drive) continuously. In contrast, in the previous event, *Terry walked for an hour, which* denotes a *walk* as a 'block' of process with one-hour duration in the past.

**Accomplishments**
(a) Terry *built five houses in two months.*

*Fig.* 22 shows the TM representation of *Terry building five houses in two months*—initial events in the figure end with the building process. The assumption is that such a process, when completed, terminates in establishing the physical body of houses. Each house is created as an entity (blueprint, etc.) and is completed physically at the end of the process of building.

The difference with activities of the walk and the driving in the previous two example is that there is a 'possibility' that walking and driving continue after finishing them while building stops at the end of the process of building. So it seems that activities either terminate or continue in the next event (e.g., After *Terry sets down to rest*) while the building is assumed to terminate in the succeeding event, i.e., is not followed by a building event.

(b) *The child is drawing a circle.*

Fig. 23 shows the diagram of *The child is drawing a circle*. The circle can be modeled as coming from the outside (e.g., instruction to draw a figure) or from memory.

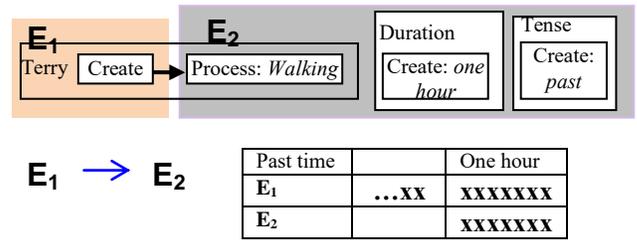

Fig. 20 TM model of *Terry walked for an hour*

| Past time | | One hour |
|---|---|---|
| E₁ | ...xx | xxxxxxx |
| E₂ | | xxxxxxx |

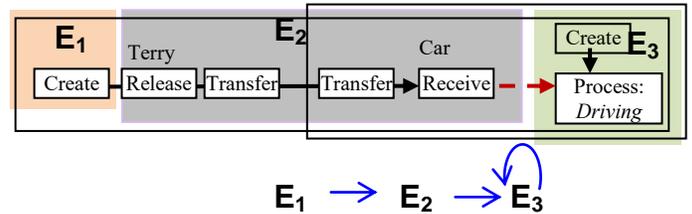

Fig. 21 TM model of *Terry is driving the car now* as an event

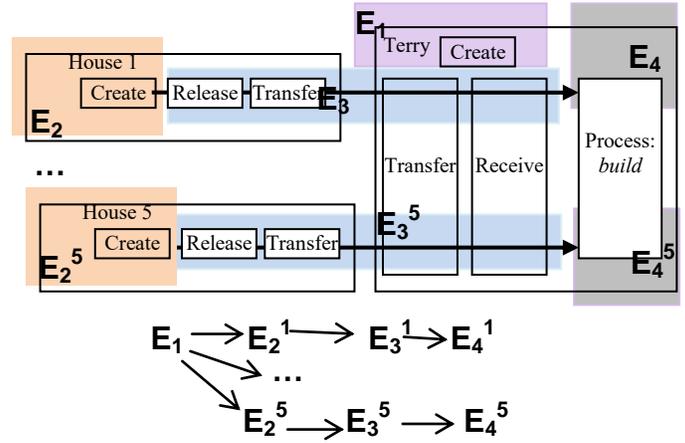

Fig. 22 TM representation of *Terry built five houses in two months*

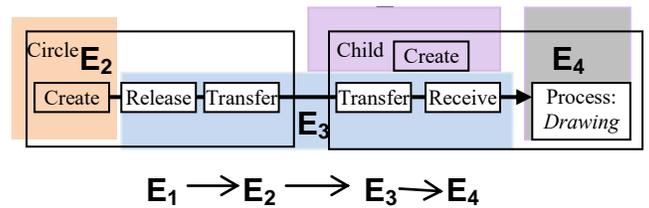

Fig. 23 TM representation of *The child is drawing a circle*.

In conclusion, the examples provided additional insight into TM modeling diverse situations. Nevertheless, while we can continue to model more examples in TM diagrams, this is unnecessary. All these classifications can be expressed in TM diagrams of various complexities, and a structure-based cataloging system can be developed based, partially, on graph theory (e.g., linear, reflexive graphs). This will be the subject of future research. Meanwhile, we can continue exploring how to model event classification using different approaches.

### B. Classifying Eventualities

[25] introduced a scheme of distinctions based on what is called eventualities. He defined eventualities as a 'special phenomenon' and proposed that they were a broader notion of events. The distinctions between different classes of eventualities are rooted in durativity (having a specific duration) and telicity (having an explicit condition of termination).

Some examples of eventualities are as follows.

(1) *John kisses Mary (atomic event)*: Fig. 24 shows the TM dynamic model of *John kissing Mary*, which is divided into four events. The term *atomic* may refer to event $E_4$, which cannot become an event without realising the other events.

(2) *Mary stumbled and twisted her ankle* (plural event): Fig. 25 shows the TM dynamic model of *Mary's stumbled and twisted ankle*. The term *plural* may refer to consecutive events $E_3$ and $E_4$.

Such events as atomic and plural seem to reflect types of dynamic graphs (serial, star, etc.). A more comprehensive approach may be developed using the TM diagrams.

### C. TimeML events

The automatic recognition of temporal and event expressions in natural language text has become an active area of research in computational linguistics and semantics [26]. *TimeML* is a specification language for the event and temporal expressions in natural language text developed in the context of question-answering systems. According to [27], TimeML defines an event as "a cover term for situations that happen or occur". Events once annotate, are classified into one of seven classes:

1. Reporting: Events where a person or organization declares, narrates or says something. Examples: say, report, announce. Fig. 26 shows the TM static model of *A person creating a report*.

2. Perception: Events that involve the physical perception of another event. Examples: see, hear. Fig. 27 shows the TM static model of *A person hears a sound*.

In this approach, other classes includes State, e.g., on-board, love, occurrence, e.g., die, crash, build, sell, I-Action: e.g., attempt, try and offer, Aspectual: e.g., begin, finish.

At this point in our discussion, we conclude that all these situations can be modelled in TM. Viewing these TM diagrams as graphs can provide an alternative event classification scheme.

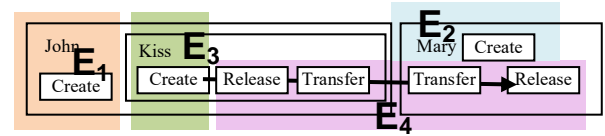

Fig. 24 TM dynamic model of *John kisses Mary*

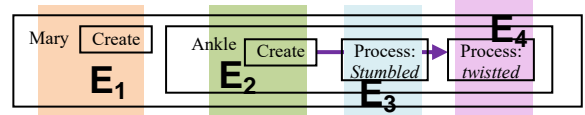

Fig. 25 TM dynamic model of *Mary stumbled and twisted her ankle*

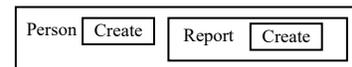

Fig. 26 TM static model of *A person creating a report*.

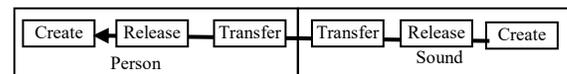

Fig. 27 TM static model of *A person hears a sound*.

### CONCLUSION

As a sequel to evolving research about a diagrammatic methodology called thinking machine (TM), this paper enhances some TM aspects related to structuring events, such as absent events. The enhancement of the TM modelling has proved to enrich conceptual modelling, hence, further research will continue in the future.

Additionally, the paper has focused on classifying events and the relationships that can be recognised among them. In conclusion, the examples of event classification have provided additional insight into TM modeling of diverse situations. Nevertheless, we can conclude that current event classification can be expressed in TM diagrams of various complexities. A structure-based cataloguing system can be developed based partially on graph theory (e.g., linear, reflexive graphs). This will be the subject of future research.